\documentclass[a4paper]{article}
\usepackage{spconf,amsmath,graphicx,cite}
\usepackage{array} % AE
\usepackage{comment,psfrag,amssymb,booktabs,tabularx,xcolor} % AE
\newcommand{\argmin}{\operatornamewithlimits{argmin}} % AE

\title{Temporal Error Concealment for Fisheye Video Sequences\\Based on Equisolid Re-Projection}
\name{Andrea Eichenseer, J\"urgen Seiler, Michel B\"atz, and Andr\'e Kaup}
\address{Multimedia Communications and Signal Processing\\
	Friedrich-Alexander University Erlangen-N\"urnberg (FAU), Cauerstr. 7, 91058 Erlangen, Germany\\
	}
\begin{document}

\maketitle
\begin{abstract}
Wide-angle video sequences obtained by fisheye cameras exhibit characteristics that may not very well comply with standard image and video processing techniques such as error concealment. This paper introduces a temporal error concealment technique designed for the inherent characteristics of equisolid fisheye video sequences by applying a re-projection into the equisolid domain after conducting part of the error concealment in the perspective domain. Combining this technique with conventional decoder motion vector estimation achieves average gains of 0.71 dB compared against pure decoder motion vector estimation for the test sequences used. Maximum gains amount to up to 2.04 dB for selected frames.
%For selected frames, maximum gains amount to 2.00 dB for the synthetically generated sequences and 1.67 dB for the real-world videos.

\end{abstract}
\section{Introduction}
\label{sec:intro}

Video surveillance, automotive, and also outdoor applications often make use of very wide fields of view (FOV) of 180 degrees and beyond.
To capture such ultra wide-angle video sequences with a single camera, fisheye lenses~\cite{miyamoto1964fel} based on projection functions quite different from the pinhole model are employed.
Many applications require the immediate coding of the obtained fisheye videos using a block-based hybrid video codec~\cite{wiegand2003avc,sullivan2012hevc}, for instance. Subsequently transmitting the coded data from the camera to a receiver over error-prone channels may cause losses that the receiver side may want to conceal to reconstruct the visual quality to a certain degree.

Countless error concealment techniques can be found in literature, classified into three categories, namely spatial, temporal, and spatio-temporal techniques.
Spatial error concealment techniques~\cite{sfg} rely only on information available in the video frame to be reconstructed. 
%and can be based on stochastic factor graphs, for instance.
%Since spatial correlations are limited, spatial error concealment 
%work better in homogeneous image areas.
Temporal error concealment approaches like decoder motion vector estimation (DMVE)~\cite{dmve} exploit correlations 
%are exploited by conducting a motion search in a reference frame 
within the temporal neighborhood of the distorted frame.
%Since translational motion can be compensated quite well, temporal error concealment techniques often lead to very good results.
The third category comprises spatio-temporal error concealment techniques which try to suitably combine spatial and temporal approaches~\cite{ifs}.
Further improvement can be achieved by adding post-processing steps like denoising~\cite{dter}.

\begin{figure}[t]
%\vspace{0.15cm}
\centering
\centerline{\includegraphics[width=\columnwidth]{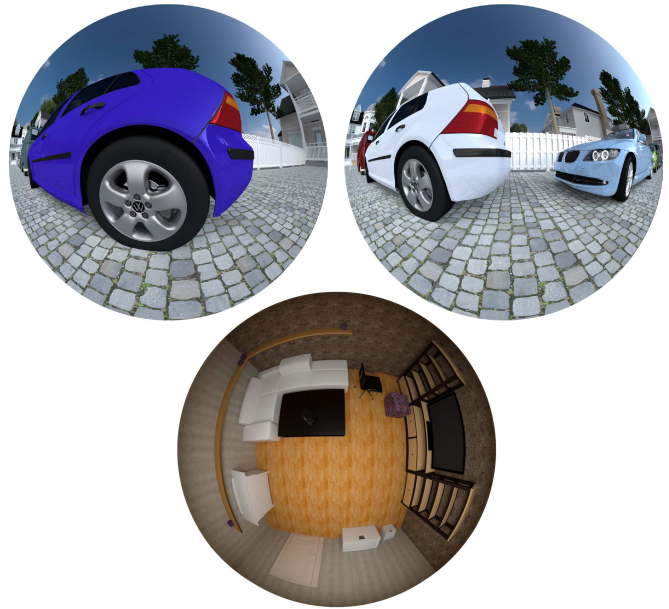}}
\vspace{-0.2cm}
\caption{\small Example frames of synthetically generated equisolid fisheye video sequences. Top: \textit{Street}, bottom: \textit{Room}.}
\vspace{-0.4cm}
\label{fig:synthvideo}
\end{figure}

In this paper, we consider temporal error concealment for fisheye video sequences.
More specifically, we propose an adapted DMVE technique designed for equisolid fisheye data as depicted in Fig.~\ref{fig:synthvideo}.
Block-matching error concealment techniques~\cite{principles} are based on a translational motion model and are thus very much suited to rectilinear video data.
For fisheye videos, however, the translational model is not a suitable assumption as they do not comply with the pinhole model.
This was partly investigated in~\cite{eichenseer2014dcorfish}, where it was shown that inter-frame video coding and, consequently, traditional motion estimation works better in the perspective domain.
As this observation can be extended towards block-matching error concealment methods like DMVE, our equisolid temporal error concealment (E-TEC) technique is based on a transform into the perspective domain to exploit its better suitability to the translational motion model.
Following the motion search in the perspective domain, E-TEC employs a re-projection into the equisolid domain, where the actual concealment is conducted.
E-TEC thus adapts the motion estimation technique described in~\cite{eichenseer2015motionfish} for use in temporal error concealment.

%Section~\ref{sec:stateoftheart} provides an outline of DMVE.
%In Section~\ref{sec:equisolid}, we present the concept of equisolid re-projection as well as our proposed technique. Section~\ref{sec:results} provides the simulation results followed by a conclusion in Section~\ref{sec:conclusion}.

\section{Decoder Motion Vector Estimation}
\label{sec:stateoftheart}

\begin{figure}[t]
\centering
\psfrag{a}[cb][rt][1.00][0]{\hspace{0.2cm}$s_{\tau-1}[m,n]$}
\psfrag{e}[cb][rt][1.00][0]{\hspace{0.2cm}$s_{\tau}[m,n]$}
\centerline{\includegraphics[width=0.85\columnwidth]{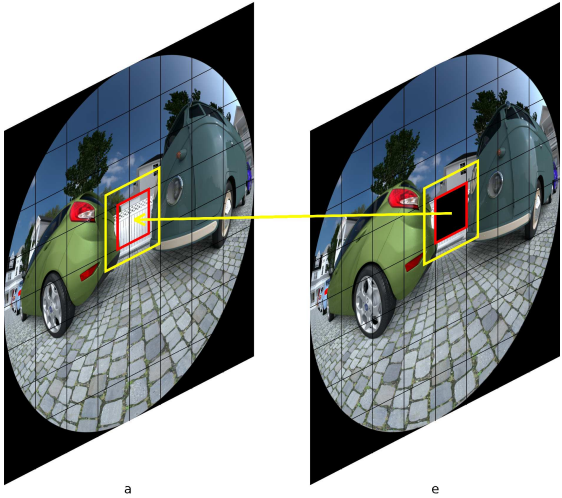}}
\vspace{-0.3cm}
\caption{\small DMVE example with decision area $\mathcal{D}$ and corresponding motion vector depicted in yellow and loss area $\mathcal{L}$ depicted in red.}
\vspace{-0.2cm}
\label{fig:dmveexmp}
\end{figure}

In the following, the principle of DMVE~\cite{dmve} is briefly outlined.
Fig.~\ref{fig:dmveexmp} visualizes the approach.
Given a video frame $s_{\tau}[m,n]$ at time $t=\tau$ containing a block loss, an error concealed frame $\tilde s_{\tau}[m,n]$ is obtained by:
\vspace{-0.2cm}
\begin{equation}
\label{eq:conceal}
\tilde s_{\tau}[m,n] = 
\begin{cases}
s_{\tau - 1}[m+\Delta m,n+\Delta n]\,, & \hspace{-0.2cm}\forall (m, n) \in \mathcal{L} \\
s_{\tau}[m,n]\,, & \hspace{-0.2cm}\forall (m, n) \notin \mathcal{L}\,.
\end{cases}
\vspace{-0.2cm}
\end{equation}
Here, $\mathcal{L}$ describes the area of the lost block and $(\Delta m, \Delta n)$ denotes the motion vector used for concealing this block.
The optimum motion vector is selected from a set of motion vector candidates $(\Delta m_i, \Delta n_i)$ defined within a certain search range and based on an error criterion such as the sum of squared differences (SSD):
% between the original and the reference block:
\begin{comment}
\begin{align}
\text{SSD}_i = \sum\limits_{(m, n)} (s_{\tau}[m,n] - s_{\tau-1}[m + \Delta m_i, n + \Delta n_i])^2 \\ 
\forall (m,n) \in \mathcal{D}_k \nonumber \:.
\end{align}
\end{comment}
\vspace{-0.2cm}
\begin{equation}
\text{SSD}_i = \hspace{-0.25cm} \sum\limits_{(m, n) \in \mathcal{D}} \hspace{-0.25cm} (s_{\tau}[m,n] - s_{\tau-1}[m + \Delta m_i, n + \Delta n_i])^2 \:.
\vspace{-0.2cm}
\end{equation}
$\mathcal{D}$ describes a decision area around the lost block, excluding the loss area $\mathcal{L}$.
In Fig.~\ref{fig:dmveexmp}, $\mathcal{D}$ comprises the area within the yellow block without the area of the red block.
Minimizing SSD$_i$ yields the motion vector to be used for concealing $\mathcal{L}$:
\vspace{-0.2cm}
\begin{equation}
(\Delta m, \Delta n)= \argmin\limits_{(\Delta m_i, \Delta n_i)} \text{SSD}_i \:.
\vspace{-0.2cm}
\end{equation}
After obtaining the motion vector, the lost block can be substituted and thereby concealed by copying the corresponding block shifted by the motion vector from the reference frame $s_{\tau-1}[m,n]$ into the distorted frame as defined in~(\ref{eq:conceal}).

Just like conventional block-based motion estimation~\cite{blockmatching} methods, DMVE relies on a translational motion model as this describes the predominant kind of motion in a typical video sequence.
Since fisheye images are not based on a perspective projection function, they exhibit characteristics for which this model no longer holds true.
We hence propose taking into account the projection function of fisheye images and introduce an adapted temporal error concealment method.
% as described below.

\section{Temporal Error Concealment via Equisolid Re-Projection}
\label{sec:equisolid}

\begin{figure}[t]
\centering
\centerline{\includegraphics[width=\columnwidth]{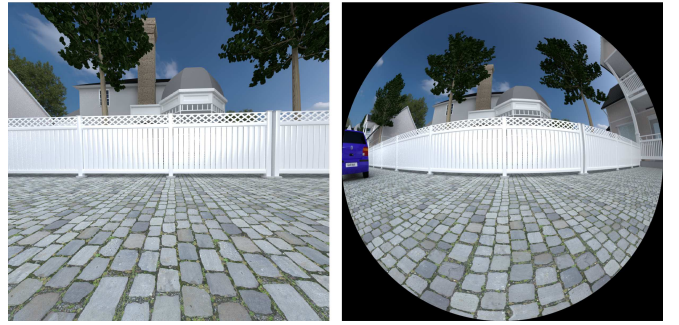}}
\vspace{-0.2cm}
\caption{\small Comparison of a perspective image (left) to its equisolid fisheye version (right) using the same spatial resolution.}
\vspace{-0.2cm}
\label{fig:rectifiedexmp}
\end{figure}

The different projection functions and resulting image characteristics of perspective images and equisolid fisheye images become quite evident by regarding~Fig.~\ref{fig:rectifiedexmp}.
While the left image is obtained using the pinhole model
\begin{equation}
r_\mathrm{p} = f \tan\theta\:,
\end{equation}
i.\,e., perspective projection, the right one is based on equisolid projection~\cite{miyamoto1964fel}: 
\begin{equation}
%\label{eq:equisolid}
r_\mathrm{e} = 2 f \sin(\theta/2)\:.
\end{equation}
In both cases, $\theta$ is the incident angle of light and $f$ denotes the focal length.
%measured against the optical axis
$r_\mathrm{p}$ and $r_\mathrm{e}$ describe the distance to the image center in the perspective and equisolid image, respectively.
%They represent the radius in the polar coordinate notations $(r_\mathrm{p},\phi_\mathrm{p})$ and $(r_\mathrm{e},\phi_\mathrm{e})$ of both the perspective and equisolid image, while $\phi_\mathrm{p}$ and $\phi_\mathrm{e}$ represent the angle.
Using polar coordinates $(r_\mathrm{p},\phi_\mathrm{p})$ and $(r_\mathrm{e},\phi_\mathrm{e})$, $r_\mathrm{p}$ and $r_\mathrm{e}$ represent the radius, while $\phi_\mathrm{p}$ and $\phi_\mathrm{e}$ denote the angle.
Evidently, equisolid projection allows a much larger FOV, but the resulting image no longer follows the rules of projective geometry and straight lines are mapped onto arcs.
As a consequence, image processing techniques based on a translational motion model must be considered suboptimal as concluded in~\cite{eichenseer2014dcorfish}.

We hence propose an equisolid temporal error concealment (E-TEC) technique based on DMVE which conducts the motion vector search in the perspective domain~\cite{eichenseer2015motionfish}.
Since projecting the entire equisolid image into the perspective domain is practically infeasible due to the vast amount of pixels this would result in, we instead manipulate the image coordinates $(r_\mathrm{e}, \phi_\mathrm{e})$ in a suitable fashion.
%To that end, w
We thus use
%\vspace{-0.1cm}
\begin{equation}
\label{eq:backward}
r_\mathrm{p} = f \tan\left(2\arcsin\left(\frac{r_\mathrm{e}}{2f}\right)\right)
%\vspace{-0.1cm}
\end{equation}
to back-project the image coordinates into the perspective domain $\mathcal{P}$.
Since the translational model holds here, the addition of the motion vector candidate $(\Delta m_i, \Delta n_i)$ is conducted in this domain using a Cartesian representation.
Afterwards, the now shifted polar coordinates $(r'_\mathrm{p}, \phi'_\mathrm{p})$ are re-projected into the equisolid domain $\mathcal{E}$ via 
%\vspace{-0.25cm}
\begin{equation}
\label{eq:forward}
r'_\mathrm{e} = 2 f \sin\left(\frac{1}{2} \arctan\left(\frac{r'_\mathrm{p}}{f} \right) \right)
%\vspace{-0.2cm}
\end{equation}
and subsequently applied to a suitably upsampled and interpolated version of the reference frame to extract the corresponding pixel values.
Note that in neither~(\ref{eq:backward}) nor~(\ref{eq:forward}) the angle is changed in any way, so that $\phi_\mathrm{p} = \phi_\mathrm{e}$ and $\phi'_\mathrm{e} = \phi'_\mathrm{p}$.
The upsampling and interpolation of the reference frame is necessary since the Cartesian coordinates corresponding to $(r'_\mathrm{e}, \phi'_\mathrm{e})$ are no longer comprised of integer values.
To preserve a certain degree of accuracy, a suitable upsampling factor must thus be chosen, e.\,g., a factor of 8 for eighth-pixel accuracy.

%In the following, we describe our proposed method in detail.
Apart from the additional projections described above, the principle of the motion vector search, including the minimization of the SSD based on a decision area $\mathcal{D}$ around the lost block, 
%, as well as the error concealment itself 
is the same as for regular DMVE.
%The main difference is the block extraction from the reference frame for the actual concealment.
To minimize the SSD of the decision area $\mathcal{D}$ and thus determine the motion vector $(\Delta m,\Delta n)$, all image coordinates $(r_\mathrm{e}, \phi_\mathrm{e}) \in \mathcal{D}$ are projected into the perspective domain, where the motion vector candidate $(\Delta m_i, \Delta n_i)$ is added.
The resulting shifted coordinates are subsequently re-projected into the equisolid domain, where they can be applied to the reference frame, thus extracting the pixel values to be compared to $\mathcal{D}$.
Repeating this for all motion vector candidates within the search range finally yields $(\Delta m,\Delta n)$.
Having determined the motion vector $(\Delta m,\Delta n)$, the block to be used for concealing the loss area is obtained by projecting all image coordinates $(r_\mathrm{e}, \phi_\mathrm{e}) \in \mathcal{L}$ into the perspective domain, adding the motion vector, and re-projecting the shifted coordinates into the equisolid domain. 
Applying the image coordinates thus obtained to the upsampled reference frame then yields the block used for concealing the area $\mathcal{L}$ of the regarded lost block.
%We not necessarily copy a block of connected pixels from the reference frame, but rather extract a set of points to be mapped onto a block to fit the concealment area.
%The red boxes in the top path of Fig.~\ref{fig:blockdiagram} depict the described equisolid re-projection approach schematically.

\begin{figure}[t]
\centering
\psfrag{c}[cb][cb][0.9][0]{$\mathbf{S}_{\tau}$}
\psfrag{r}[ct][ct][0.9][0]{$\mathbf{S}_{\tau-1}$}
\psfrag{e}[ct][ct][0.9][0]{$\mathbf{\tilde S}_{\tau}$}
\psfrag{S}[cc][cB][0.9][0]{SSD}
\psfrag{o}[cc][cB][0.9][0]{min}
\psfrag{b}[cc][cB][0.9][0]{traditional DMVE}
\psfrag{x}[cc][cB][0.9][0]{\shortstack{temporal error concealment\\via equisolid re-projection}}
\psfrag{P}[cc][Bc][1.2][0]{$\mathcal{P}$}
\psfrag{E}[cc][Bc][1.2][0]{$\mathcal{E}$}
\psfrag{M}[cB][cB][0.85][0]{$(\Delta m_i, \Delta n_i)$} %MV candidate}
\centerline{\includegraphics[width=\columnwidth]{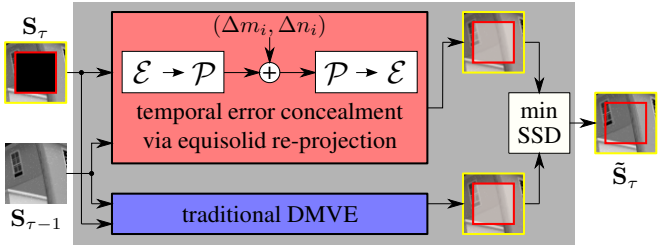}}
\vspace{-0.3cm}
\caption{\small Schematic depiction of HE-TEC (gray) combining the proposed E-TEC (light red) with conventional DMVE (blue).}
\vspace{-0.4cm}
\label{fig:blockdiagram}
\end{figure}

To implement the presented E-TEC method, we directly build upon conventional DMVE and thereby create a hybrid equisolid temporal error concealment (HE-TEC) technique.
HE-TEC allows DMVE as an optional technique for the concealment of blocks where our E-TEC method meets its limits.
One such limiting factor is the inverse tangent function in~(\ref{eq:forward}).
%As the set of motion vector candidates to be tested is, in our case, defined within a given integer-pixel search range, conducting the actual addition in the perspective domain leads to the search area describing an integer grid.
As we define an integer-pixel search range in the perspective domain, employing~(\ref{eq:forward}) leads to a shortened search range in the equisolid domain.
%This is due to $r'_\mathrm{e}$ being located closer to the image center than its corresponding $r'_\mathrm{p}$.
%In other words,
Since the original search range in the perspective domain is able to cover a larger range of motion, DMVE may outperform E-TEC in the case of fast motion or for lost blocks in the periphery of the fisheye image, as it is not inhibited by a shortened search range.
This is especially true when nearing the 180 degree boundary of the fisheye image as these coordinates are located near infinity in the perspective domain.
Any search range is consequently re-projected onto a very small area or even a single point in the equisolid domain according to the inverse tangent in (\ref{eq:forward}), hence being unable to capture any kind of motion between frames.
%, whereas our proposed method may prove advantageous for the center of the image due to its implicit sub-pixel accuracy.

For the proposed HE-TEC, we thus incorporate an SSD-based decision between pure DMVE and our equisolid re-projection variant E-TEC.
HE-TEC is schematically depicted in Fig.~\ref{fig:blockdiagram}, where $\mathbf{S}_{\tau}$, $\mathbf{S}_{\tau-1}$, and $\mathbf{\tilde S}_{\tau}$ denote the lossy signal to be concealed, the reference frame, and the error concealed signal, respectively.
The blue box denotes the conventional DMVE approach.
The light red box describes the proposed equisolid re-projection approach E-TEC. 
While only the projection of the image coordinates into the perspective domain $\mathcal{P}$, the motion vector addition, and the re-projection of the translated image coordinates into the equisolid domain $\mathcal{E}$ are explicitly visualized, any other necessary processing steps like the motion vector search based on SSD minimization, the upsampling of the reference frame, as well as the concealment block extraction are also part of E-TEC.
%the extraction of the corresponding pixel values from the upsampled reference frame.
%Note that only for the extraction step, the reference frame must be available.
Note that the projections do not require any information about actual pixel values as they work solely with image coordinates.
In the following, HE-TEC is compared against pure DMVE.

\section{Simulation Setup and Results}
\label{sec:results}

\begin{figure}[t]
\centering
\centerline{\includegraphics[width=\columnwidth]{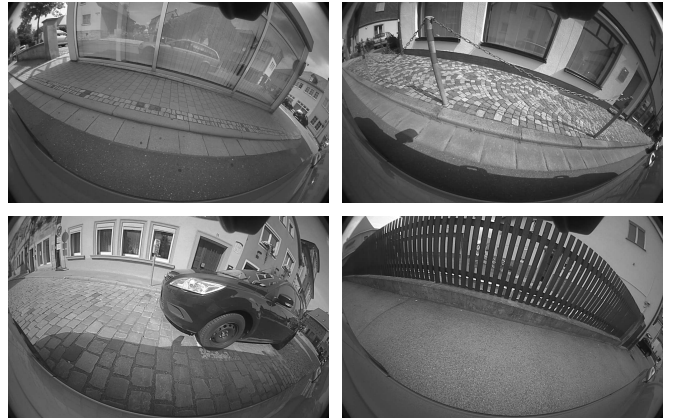}}
\vspace{-0.3cm}
\caption{\small Example frames of real-world fisheye video sequences. Top left to bottom right: \textit{Video1}, \textit{Video2}, \textit{Video3}, and \textit{Video4}.}
\vspace{-0.2cm}
\label{fig:realvideo}
\end{figure}

\begin{table}[t]
\vspace{0.2cm}
%\label{tab:setup}
\small
\centering
\renewcommand\arraystretch{0.9}
\begin{tabularx}{\columnwidth}{ccccc}
\toprule
\textbf{Sequence} & \textbf{Resolution} & \textbf{Frames} & \textbf{Length} & \textbf{Frames}\\
\textbf{name} & \textbf{(pixels)} & \textbf{per second} & \textbf{(frames)} & \textbf{tested} \\
\midrule
\textit{Street} & 1088$\times$1088 & 25 & 1800 & 55 \\ %frameno =[100:50:350,400:10:700,800:50:1650]
\textit{Room} & 1088$\times$1088 & 25 & 400 & 7 \\
\addlinespace
\textit{Video1} & 768$\times$1216 & 30 & 30 & 29 \\
\textit{Video2} & 768$\times$1216 & 15 & 30 & 29 \\
\textit{Video3} & 768$\times$1216 & 30 & 30 & 29 \\
\textit{Video4} & 768$\times$1216 & 30 & 30 & 29 \\
\bottomrule
\end{tabularx}
\vspace{-0.35cm}
\caption{\small General information on the test sequences.}
\vspace{-0.4cm}
\label{tab:setup}
\end{table}

\begin{table}[t]
\vspace{0.3cm}
\small
\centering
\renewcommand\arraystretch{0.9}
\begin{tabularx}{\columnwidth}{p{1.4cm}cccp{0.2cm}c}
\toprule
& DMVE & HE-TEC & Gain & & Gain$_{\text{max}}$ \\
\midrule
\textit{Street} & 29.62 dB & 30.69 dB & \textbf{1.07 dB} & & 2.04 dB \\
\textit{Room} & 44.64 dB & 45.06 dB & \textbf{0.42 dB} & & 0.78 dB \\
\addlinespace
\textit{Video1} & 30.96 dB & 31.62 dB & \textbf{0.66 dB} & & 1.48 dB \\
\textit{Video2} & 24.76 dB & 25.47 dB & \textbf{0.71 dB} & & 1.03 dB \\
\textit{Video3} & 28.53 dB & 29.18 dB & \textbf{0.65 dB} & & 0.87 dB \\
\textit{Video4} & 28.88 dB & 27.65 dB & \textbf{0.77 dB} & & 1.67 dB \\
\midrule
Average & & & 0.71 dB & & \\ 
\bottomrule
\end{tabularx}
\vspace{-0.35cm}
\caption{\small Average luminance PSNR results. In addition, the overall maximum gain achieved for a selected frame is given.}
\vspace{-0.35cm}
\label{tab:psnr}
\end{table}

\begin{figure*}[t]
\centering
\psfrag{a}[Bc][Bc][.9]{(a)}
\psfrag{b}[Bc][Bc][.9]{(b)}
\psfrag{c}[Bc][Bc][.9]{(c)}
\psfrag{d}[Bc][Bc][.9]{(d)}
\psfrag{e}[Bc][Bc][.9]{(e)}
\centerline{\includegraphics[width=\textwidth]{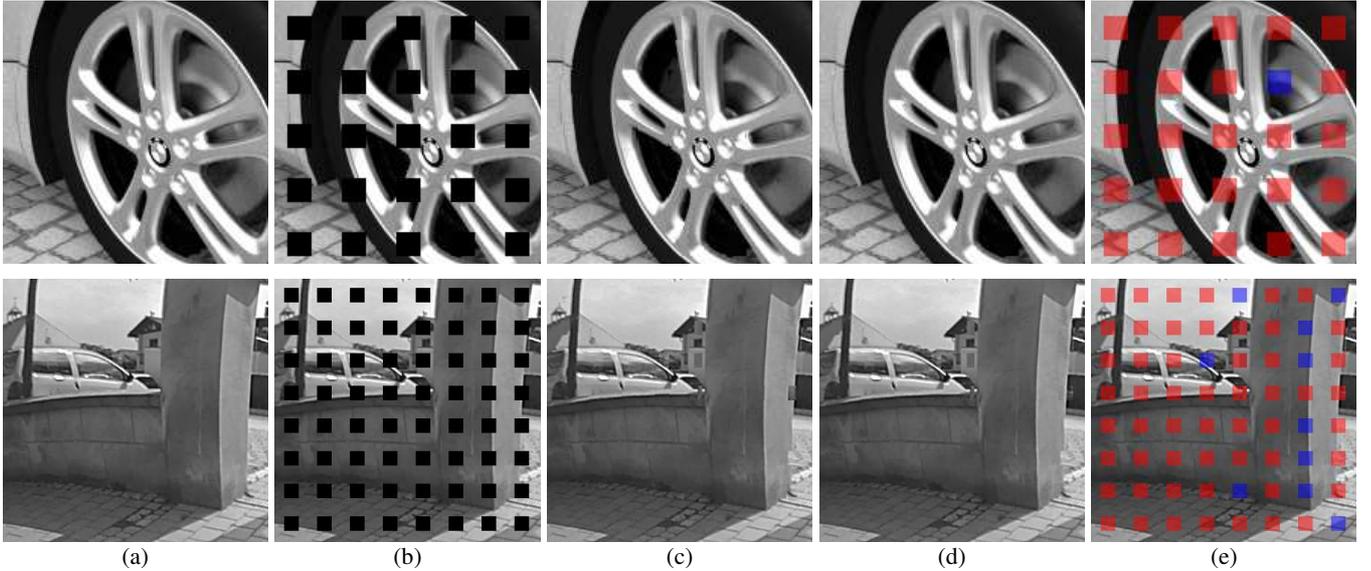}}
\vspace{-0.35cm}
\caption{\small Exemplary image detail results for \textit{Street} (top) and \textit{Video1} (bottom). (a) original image, (b) lossy image, (c) pure DMVE, (d) HE-TEC, (e) HE-TEC overlaid with red when the E-TEC approach was given preference and blue when conventional DMVE was performed.}
\vspace{-0.5cm}
\label{fig:resexmp}
\end{figure*}

To test our HE-TEC approach, we generated synthetic fisheye video sequences using blender~\cite{blender} and, to that end, made use of several object models from~\cite{blendswap} to create realistic scenes.
The blender setting for the camera was panoramic fisheye using equisolid projection.
The FOV was set to 185 degrees, the focal length to 1.8 mm, and the sensor size to 5.2 mm by 5.2 mm, so that the entire circular fisheye can be captured.
Fig.~\ref{fig:synthvideo} shows exemplary frames of our synthetic fisheye sequences.
\textit{Street} employs a moving camera and static objects and the contained motion is mostly translational. \textit{Room} on the other hand uses a static camera and various moving objects so that there is no global motion.

To further test HE-TEC on real-world video sequences, we used four traffic sequences which all contain global translational motion.
For each real-world sequence, an example frame is depicted in Fig.~\ref{fig:realvideo}.
Regarding the FOV and focal length, we assume the same values as used for the synthetic sequences.
Only the sensor size was changed to 4.6 mm by 2.9 mm, as the real-world sequences evidently consist of full-frame fisheye images which fill the entire sensor area.
The sensor size was estimated by searching the maximum radius that was mapped onto the image plane along with the assumption that 5.2 mm is enough to represent the entire circular fisheye.
Further information on the test sequences used is given in Table~\ref{tab:setup}.

In all of the tests conducted, a fixed integer-pixel search range of 128 pixels in every direction was used for both regular DMVE as well as our HE-TEC technique so that 257$\times$257 motion vector candidates were evaluated for each lost block.
For HE-TEC, the re-projected image coordinates were applied to a reference frame upsampled by a factor of 8 using cubic convolution interpolation.
The proposed HE-TEC technique was evaluated for multiple isolated block losses of 16$\times$16 pixels throughout all tests conducted.
The decision area $\mathcal{D}$ around each lost block was set to a width of 8 pixels so that the union $\mathcal{D} \cup \mathcal{L}$ forms an area of 32$\times$32 pixels.
% and evaluated at 100 frame intervals of \textit{Street}, 150 frame intervals of \textit{Room}, as well as for 30 consecutive frames each of \textit{Video1} through \textit{Video4}.

Table~\ref{tab:psnr} summarizes the average luminance PSNR results calculated for the loss areas as well as the average gains obtained for each sequence.
Additionally, the maximum gains achieved for selected frames of each sequence are given.
For \textit{Street}, average gains amount to 1.07 dB with an overall maximum of 2.04 dB.
This result shows that equisolid re-projection is a suitable means for an improved motion search in fisheye sequences.
Not surprisingly, there is a much lower gain for the static camera sequence \textit{Room} since DMVE is able to achieve perfect signal reconstruction for most of the lost blocks and HE-TEC cannot improve on that.
Nonetheless, small gains can be achieved if lost blocks contain parts of the few moving objects within this sequence.

In terms of real-world fisheye sequences, average gains of around 0.7 dB are obtained, showing that the proposed HE-TEC technique also works on non-synthetic sequences.
Although the assumption of an equisolid projection is certainly not an accurate one, it is good enough to achieve an improved concealment result.
Adapting HE-TEC to calibrated projection information should further increase the gain.
As mentioned earlier, the search range is still a limiting factor of HE-TEC, so that an adaptation considering the equisolid projection should also potentially increase the obtained gains.

%A visual example showing the error pattern as well as the obtained DMVE and HE-TEC concealment results for part of a frame is given in Fig.~\ref{fig:resexmp}.
A zoomed-in visual example is given in Fig.~\ref{fig:resexmp}.
(b) shows the error pattern, (c) the concealment results obtained by DMVE, and (d) the HE-TEC results. 
In (e), the HE-TEC results are additionally overlaid with a decision mask.
Red color denotes those lost blocks for which error concealment was done using the projection-based E-TEC approach, while blue color denotes blocks for which conventional DMVE was chosen.
Although the overall visual impression seems very similar for both DMVE and HE-TEC, differences along curved shapes can be made out upon closer inspection.
Here, the equisolid re-projection technique is able to achieve better reconstruction results that sum up to an improved image quality.
These visual results are representative for all frames tested, synthetic and real-world data alike.

When evaluating the HE-TEC results for the real-world sequences, it was observed that on average, 75\,\% of all lost blocks were concealed via E-TEC, i.\,e., via equisolid re-projection, while conventional DMVE was chosen for only 25\,\% of the blocks.
For \textit{Street} and \textit{Room}, E-TEC was employed for 71\,\% and 96\,\% of all lost blocks, respectively.
It is quite evident that our equisolid re-projection technique is chosen for most of the lost blocks, thus substantiating the PSNR results.
Since the majority of lost blocks is concealed by the introduced E-TEC approach, its implementation as a stand-alone error concealment method is also conceivable, rendering the blue box as well as the SSD-based decision in Fig.~\ref{fig:blockdiagram} obsolete.

%\vspace{0.5cm}
\section{Conclusion}
\label{sec:conclusion}

In this paper, we introduced a temporal error concealment technique for fisheye video sequences via equisolid re-projection.
Based on the knowledge that the translational motion model does not hold for fisheye videos due to the different underlying projection function, we employed suitable projections from the equisolid to the perspective domain and vice versa in order to conceal the lost blocks with the help of a reference frame.
Furthermore, a hybrid technique combining this approach with conventional DMVE was proposed and evaluated.
Average gains in luminance PSNR amounted to 
%a maximum of 2.00 dB for a selected frame of \textit{Street} and 1.67 dB for a selected frame of \textit{Video4}.
%On average, gains of 
0.71 dB for both the synthetic and real-world sequences tested, letting us conclude that exploiting knowledge about optics that differ from the conventional pinhole model is a suitable means for improving image reconstruction quality.

The development of the proposed E-TEC method as a stand-alone technique is part of work in progress.
A major point of interest to that end is the suitable handling of the peripheral parts of circular fisheye frames, i.\,e., those parts, where the FOV gets close to and surpasses the 180 degree boundary.
Current work also investigates spatio-temporal error concealment for fisheye video sequences as well as optimizations with regard to the motion search.
Of particular interest here is the reduction of motion vector candidates to evaluate.
%Future work will furthermore take into consideration denoising as a refinement step for spatio-temporal error concealment as well as error concealment for single fisheye images.

%Wie wichtig ist eine realistische Anwendung? (FMO etc.)
%Macht eine Variation der Blockgroesse Sinn?

\vfill\pagebreak
\section{Acknowledgment}
%\vspace{-0.2cm}
%\small
This work was partly supported by the Research Training Group 1773 “Heterogeneous Image Systems”, funded by the German Research Foundation (DFG).
The real-world fisheye sequences were kindly provided by Continental Chassis \& Safety  BU ADAS Segment Surround View (A.D.C GmbH), Kronach.
\vspace{-0.4cm}

\bibliographystyle{IEEEbib}
\bibliography{refs}

\end{document}